\documentclass[twocolumn, tighten, times, 12pt]{aastex62}
\usepackage{latexsym, graphicx, amssymb, longtable, epsf,color,amsmath,amssymb}
\usepackage{comment}
\usepackage{natbib}
\def\etal{{et~al.}}

\def\HII{{H\,{\textsc{ii}}}}
\def\Msun{{M$_\odot$}}
\def\vvcrit{{$v/v_{\rm crit}$}}

\hypersetup{linkcolor=red,citecolor=blue,filecolor=blue,urlcolor=magenta}

\shorttitle{Stromlo Star tracks}
\shortauthors{Grasha \etal}

\begin{document}
\title{Stromlo Stellar Tracks: non-solar scaled abundances for massive stars}

\author[0000-0002-3247-5321]{K. Grasha}
\email{kathryn.grasha@anu.edu.au}
\affiliation{Research School of Astronomy and Astrophysics, Australian National University, Canberra, ACT 2611, Australia}   
\affiliation{ARC Centre of Excellence for All Sky Astrophysics in 3 Dimensions (ASTRO 3D), Australia}   

\author[0000-0002-5021-6737]{A. Roy}
\affiliation{Research School of Astronomy and Astrophysics, Australian National University, Canberra, ACT 2611, Australia}   
\affiliation{ARC Centre of Excellence for All Sky Astrophysics in 3 Dimensions (ASTRO 3D), Australia}   

\author[0000-0002-6620-7421]{R. S. Sutherland}
\affiliation{Research School of Astronomy and Astrophysics, Australian National University, Canberra, ACT 2611, Australia}   
\affiliation{ARC Centre of Excellence for All Sky Astrophysics in 3 Dimensions (ASTRO 3D), Australia}   

\author[0000-0001-8152-3943]{L. J. Kewley}
\affiliation{Research School of Astronomy and Astrophysics, Australian National University, Canberra, ACT 2611, Australia}   
\affiliation{ARC Centre of Excellence for All Sky Astrophysics in 3 Dimensions (ASTRO 3D), Australia}

\begin{abstract}
We present the Stromlo Stellar Tracks, a set of stellar evolutionary tracks, computed by modifying the Modules for Experiments in Stellar Astrophysics (MESA) 1D stellar evolution package, to fit the Galactic Concordance abundances for hot ($\mathrm{T} > 8000$ K) massive ($\geq 10$~\Msun) Main-Sequence (MS) stars. Until now, all stellar evolution tracks are computed at solar, scaled-solar, or alpha-element enhanced abundances, and none of these models correctly represent the Galactic Concordance abundances at different metallicities. This paper is the first implementation of Galactic Concordance abundances to the stellar evolution models. The Stromlo tracks cover massive stars ($10\leq M/M_\odot \leq 300$) with varying rotations ($v/v_{\rm crit} = 0.0, 0.2, 0.4$) and a finely sampled grid of metallicities ($-2.0 \leq {\rm [Z/H]} \leq +0.5$; $\Delta {\rm [Z/H]} = 0.1$) evolved from the pre-main sequence to the end of $^{12}$Carbon burning. We find that the implementation of Galactic Concordance abundances is critical for the evolution of main-sequence, massive hot stars in order to estimate accurate stellar outputs (L, T$_{\rm eff}$, $g$), which, in turn, have a significant impact on determining the ionizing photon luminosity budgets. We additionally support prior findings of the importance that rotation plays on the evolution of massive stars and their ionizing budget. The evolutionary tracks for our Galactic Concordance abundance scaling provide a more empirically motivated approach than simple uniform abundance scaling with metallicity for the analysis of \HII\ regions and have considerable implications in determining nebular emission lines and metallicity. Therefore, it is important to refine the existing stellar evolutionary models for comprehensive high-redshift extragalactic studies. The Stromlo tracks are publicly available to the astronomical community online.
\end{abstract}

\section{Introduction}\label{sec:intro}
Our understanding of stellar physics and our ability to create realistic and accurate stellar evolution models across a wide range of stellar parameters heavily impact our ability to create physically realistic stellar population synthesis and photoionization models to interpret galaxy spectra. Most stellar evolutionary models (e.g., BaSTI \citep{pietrinferni04, hidalgo18}; Geneva \citep{ekstrom12}; Padova \citep{girardi02}; Y$^2$ \citep{yi01, yi03, demarque04}) assume solar or  scaled-solar abundances \citep{anders89, asplund09}. Two immediate problems arise from currently available stellar tracks. 

First, solar \citep{anders89, asplund09} relative abundance ratios conflict observed abundance ratios in \HII\ regions in the Milky Way, Large Magellanic Cloud, or Small Magellanic Cloud at a given metallicity \citep{morel09, nieva12, nicholls17}. The solar-scaled and alpha-enhanced abundances predict non-realistic stellar quantity measurements (L, $g$, T$_{\rm eff}$, etc) and thus will predict non-realistic ionizing photon budgets when used to interpret observations. This is critically important as solar-scaled stellar evolutionary tracks are in general not well matched to stellar observations and subsequent nebular emission modeling \citep[see][]{przybilla08, morel09, nicholls17, cazorla17, kewley19_araa}. The implementation of abundance patterns is especially critical in massive, hot stars that dominate the excitation sources for \HII\ regions. This is especially important because nearby OB stars quite often exhibit metal abundances that are generally lower than solar estimates \citep{morel09}.

Iron abundances relative to $\alpha$-element abundances change as a function of time and with the galactic environment \citep{wyse93} and this systematic variation of [$\alpha$/Fe] needs to be taken into account at different metallicities. It is not just iron that changes with metallicity -- a complete census on the chemical history of other elements and their evolution with overall metallicity is critical to determine accurate metallicity measurements in galaxies, especially at high redshift. The [NII]/[OII] ratio is an ideal abundance diagnostic \citep{kewley02}, but at high redshift the [OII]$\lambda\lambda$3727, 3729 doublet is often unobservable, relying on calibrations based on [NII]/H$\alpha$ \citep{denicolo02} and/or the [NII]/[OIII] ratio \citep{pettini04}. Often yet, only red line ratios are available such as H$\alpha$, [NII], and [SII], and the O/H ratio relies on indirect methods using combinations of these line ratios \citep{dopita16}. 

Second, coarse metallicity grids are usually assumed in the calculation of stellar evolutionary tracks. These coarse metallicity griding results in non-accurate values while interpolating different stellar quantities (L, $g$, T$_{\rm eff}$, etc) for the metallicities falling in between the grid metallicities. The coarse metallicity grids limit the resolution at which theoretical strong-line metallicity diagnostics can be reliably calculated \citep{kewley19_apj, kewley19_araa}. Interpolation of metallicity diagnostics in coarse metallicity grids prevents accurate measurements of nebular emission lines for star-forming galaxies and potentially has important consequences for the interpretation of galactic properties of high-redshift galaxies such as their star formation rates \citep[SFRs; e.g.,][]{kewley04} and ionization parameters which show metallicity dependence in ionization parameter diagnostics \citep{kewley02, kewley19_araa}.  

Accurate determination of the chemical evolutionary state of distant galaxies will be critically important for next generation observatories such as JWST that will reveal the first galaxies. These objects will not be physically represented with current stellar evolutionary models at Solar or alpha-enhanced abundances and will predict non-realistic stellar quantities and thus, ionizing photon budgets. New stellar evolution models that are not limited to Solar or alpha-enhanced abundance ratios are critically needed. New stellar evolution models and their opacities additionally need to be calculated at much finer metallicity intervals of $\sim$0.1--0.2 dex in log(O/H) to avoid non-accurate values while interpolating different stellar quantities. These new tracks will allow for consistent abundance ratios to be used in stellar population synthesis and photoionization models to derive accurate, high-resolution metallicity diagnostics for the first time.

This paper will enable the future development of atmosphere and photoionization nebular modeling with the same physical inputs to self-consistently predict the emission spectra arising from ionized nebula and central ionizing stellar population. This paper is the first in this series and presents the stellar evolutionary tracks using Galactic Concordance abundances to be used in modern stellar population synthesis, atmosphere, and photoionization models. The Galactic Concordance reference standard and scaling system serves as an empirical basis for interpreting observations and determining the physical conditions of \HII\ regions. 

The paper is organized as follows. In Section~\ref{sec:method}, we describe the numerical methods adopted in our stellar evolution models. Section~\ref{sec:results} reports the results and comparison between the Stromlo non-uniform abundance stellar tracks and the scaled-solar MIST stellar tracks. We discuss the impact of the abundance ratio implementation on the photon ionizing budget in Section~\ref{sec:discussion}. We conclude and summarize our results in Section~\ref{sec:conclusions}.

\section{Method: stellar evolution calculation}\label{sec:method}
In order to explore the impact of non-uniform scaled abundances on stellar evolutionary tracks, we construct the Stromlo Stellar Tracks\footnote{The Stromlo Stellar Tracks are publicly available online at \url{https://sites.google.com/view/stromlotracks}}, self-consistent stellar evolution models using the Modules for Experiments in Stellar Astrophysics \citep[MESA\footnote{\url{http://mesa.sourceforge.net}};][]{paxton11, paxton13, paxton15, paxton18, paxton19} stellar evolution code. We build upon the stellar evolutionary models used in the MESA Isochrones and Stellar Tracks (MIST) by \citet{choi16} to present models with the scaling abundances based on Milky Way stellar abundance data referred to as `Galactic Concordance' \citep{nicholls17}. We focus on massive ($>$10~$M_\odot$) hot stars that dominate the ionizing budget that power \HII\ regions and provides the feedback that regulates the efficiency of star formation \citep{mckee07_araa, krumholz12_etal, hopkins12_etal} and drives turbulence and wind outflows. 

We configure MESA to include all the same physical processes and parameters values as used in the MIST library as described by \citet{choi16} with modifications to improve the treatment of massive stars as outline by \citet{roy20}, which includes our inclusion of Galactic Concordance abundances to replace replace commonly adopted uniform scaling abundances (Section~\ref{sec:abundances}), radiative opacities (Section~\ref{sec:opacity}), and our treatment of mixing mechanisms (Section~\ref{sec:mixing}). The software used to generate the  Stromlo Tracks is the GALCON-HOT-MIST package, to be described in Roy et al. (in preparation).

For this work, we use MESA version v9793 compiled with GNU Fortran version 7.2.0 installed as part of MESA SDK\footnote{\url{http://www.astro.wisc.edu/~townsend/static.php?ref=mesasdk}}. All of our MESA calculations implement the same spatial and temporal resolution conditions as adopted by \citet{choi16} for MIST. Below we briefly outline the parameters in MESA/MIST immediately pertinent to this work and we refer the reader to \citet{paxton11, paxton13, paxton15} for more detailed information on MESA and \citet{choi16} for more detailed information regarding the physical processes in MIST. For detailed information regarding the impact of our different setups to the MIST models for high mass stars and associated uncertainties in various parameter choices and their effects, we refer the reader to \citet{roy20}.

\subsection{Non-solar elemental abundances}\label{sec:abundances}
We adopt the non-solar abundance standard developed by \citet{nieva12} and \citet{nicholls17} referred to as `Galactic Concordance' abundances. Galactic concordance is based on the observed metallicities of 29 main-sequence B-stars in local galactic region \citep{nieva12} and is augmented with elements that are of minor importance in nebular and stellar modeling \citep{nicholls17}. Galactic Concordance abundances are representative of present day, nearby massive stars \citep{nieva12} and we adopt the scaling method in order to scale elements consistently down to [Fe/H] = $-2$ using the scaling relation of \citet{nicholls17}. The Galactic Concordance reference standard and scaling method provides a reliable present-day cosmic abundance reference points for anchoring chemical evolution models to observation. Most importantly for this work and future work on creating self-consistent atmospheres and nebular models based on the same abundance, Galactic Concordance allows for linking the stellar abundance scale to the nebular abundance scale. As elements have been observed to vary systematically with [Fe/H], unlike MIST, we do not assume that all elements uniformly scale with Fe and we adopt the non-uniform scaling relation for the abundances of \citet{nicholls17}. 

We adopt the same linear fits for the elemental scaling parameter given in \citet{nicholls17} and use stellar abundance measurements using iron as the reference scale in all our models. The piecewise linear fit for iron-based scaling for the $\alpha$ and $\alpha$-like elements X is calculated as follows:
\begin{align}\label{eq:fit}
[\rm {X/Fe}] &= +\Xi_{\rm Fe} \quad  &-2.5< [\rm{Fe/H}] < -1.0 \nonumber\\
[\rm {X/Fe}] &= -\Xi_{\rm Fe} \times [\rm{Fe/H}]   \quad &-1.0< [\rm{Fe/H}] < +0.5  \nonumber \\
[\rm {X/Fe}] &= -\Xi_{\rm Fe} \times 0.5   \quad &[\rm{Fe/H}] > +0.5
\end{align}
where $\Xi_{\rm Fe}$ is the iron-based scaling factor for each element, listed in Table~2 of \citet{nicholls17}. The elements H, He, Li, Be and B are not described by the $\Xi$ parameter, because hydrogen is the reference element, we assume helium scales simply with the oxygen scaling factor. Carbon and nitrogen are not well described by a simple piecewise linear fit because of the complexities of primary and secondary enrichment and are fit via the expression:
\begin{equation}
\log(\rm{X/O}) = \log \left( 10^a + 10^{[\log(\rm{O/H})+b]} \right),
\end{equation}
where for carbon a = $-$0.8, b = 2.72, and for nitrogen a = $-$1.732, b = 2.19. For fluorine, chlorine, neon and argon, there are no extensive stellar abundance scaling data and we assume their abundances scale with oxygen.

\subsection{Radiative opacities}\label{sec:opacity}
The radiative opacity tables implemented in MESA are divided into two temperature regimes and are treated separately, high ($\log T/K >$ 4) and low ($\log T/K <$ 4) temperatures. Our model grids are all computed in the high mass regime $M>10$~\Msun\ and thus we only use the high temperature opacity tables. The radiative opacities implemented in MESA for the high temperature regime are from OPAL \citep{rogers92, iglesias93, iglesias96} or OP \citep{seaton05}. Following \citet{choi16}, we use OPAL opacities. The OPAL opacity tables in MESA and as incorporated in MIST are computed using \citet{asplund09} photospheric abundances. 

The Galactic Concordance non-uniform parametric scaling provides a more physically realistic approach than simple uniform abundance scaling with metallicity. For self-consistency in our stellar tracks, we calculate new OPAL\footnote{\url{https://opalopacity.llnl.gov}} opacity tables for our tracks that use Galactic Concordance abundances. The OPAL opacity tables must be re-computed at each metallicity value in our stellar grid (Section~\ref{sec:grid}) because we cannot assume uniform scaling as is traditionally done with the OPAL tables in MESA that assume solar abundances.

\subsection{Mixing processes}\label{sec:mixing}
In stellar evolution codes, mixing describes the convective transport of energy within the stellar interior. We implement the Ledoux criterion for the convective mixing of elements as adopted in \citet{choi16}. The only change in our mixing methods is the inclusion of the instability caused by magnetic torques by dynamo-generated fields referred to as Spruit Tayler (ST), which follows the standard approach used for high-mass stellar evolution in MESA \citep{heger00, heger05}. This is combined with the five rotationally induced instabilities MIST used: dynamical shear instability (DSI), secular shear instability (SSI), Solberg-H{\o}iland (SH) instability, Eddington-Sweet (ES) circulation, and Goldreich-Schubert-Fricke (GSF) instability. 

The diffusion coefficients for these six rotational mixing processes are combined with the diffusion coefficients for non-rotational processes: convection, convection overshoot, semiconvective mixing, and thermohaline mising. The total sum of the angular momentum and abundance diffusion equations describe the diffusion coefficients. MESA implements mixing using the common approach of treating the chemical composition $D$ and angular momentum $\nu$ transport in a diffusion approximation \citep[e.g., ][]{potter12}. We calculate $D$ and $\nu$ following \citet{choi16} with the additional inclusion ST instabilities as implemented in \citet{roy20} for high mass stellar evolution.

For further details of mixing mechanisms and angular momentum transport in MIST, we refer readers to \citet{choi16}.

\subsection{Mass loss}\label{sec:massloss}
Mass loss is one of the dominant uncertainties in evolutionary models of massive stars \citep{smith14_araa}. Our treatment of mass loss via stellar winds is based on the ``Dutch'' mass loss recipe that is standard in MESA. We adopt the \citet{vink01} mass loss prescription for metallicity-dependent winds in hot $T>10^4$~K stars. Mass loss is enhanced by rotation (Section~\ref{sec:grid}) and our prescriptions for mass loss matches the stellar wind recipe of \citet{choi16} for MIST. 

As discussed in \citet{roy20}, varying the mass-loss rate by a constant factor has minimal impact on the He surface fraction for rapidly rotating stars. This is due to rotational mixing providing a mechanism to bring He to the surface that is independent of mass-loss. On the other hand, for non-rotating stars, varying the mass-loss rate by a constant factor has a non-negligible impact on the enhancement of surface abundances. \citet{roy20} found that tripling the mass-loss rate for a 100~\Msun\ star allows surface He enrichment to occur for metallicites as low as [Fe/H] = $-1$. Reducing the mass-loss rate by a factor of 3 prevents the surface He mass fraction from ever reaching above 30\%, eliminating the WR evolutionary phase, even at solar metallicity. The main effect of increasing or decreasing the mass-loss rate thus changes the maximum amount of surface He enrichment, with a larger impact on non-rotating stars.

\subsection{Model grid}\label{sec:grid}
We calculate extensive grids of stellar evolutionary tracks that cover a wide range in stellar mass, rotation, and metallicities as follows:

\textbf{Stellar mass:} The stellar mass of evolutionary tracks ranges from 10 to 300~M$_\odot$ following the same spacing as MIST for a total of 55 models in our mass range: $\Delta M = 1 M_\odot$ in the range 10--20~M$_\odot$, $\Delta M = 2 M_\odot$ in the range 22--40~\Msun, $\Delta M = 5 M_\odot$ in the range 45--150~\Msun, and $\Delta M = 25 M_\odot$ in the range 175--300~\Msun. We choose these masses to provide sufficient coverage of the range of masses that dominate the photoionization budget for ease of input into atmosphere and nebular modeling. The models are evolved through the end of carbon burning and are stopped when the central $^{12}$C abundance drops to $10^{-4}$. 

\textbf{Stellar abundance:} We calculate grids with metallicity values from [Fe /H] = $-$2.0 to +0.5, with 0.1 dex spacing. We calculate Galactic Concordance non-solar-scaled abundance \citep{nicholls17} grids where the relative fraction of each elemental abundance as a function of metallicity is calculated according to Eqn.~\ref{eq:fit}. We also calculate protosolar abundance \citep{asplund09} grids as implemented in MIST. Following \citet{choi16}, we calculate the initial helium abundance adopting a scaling of $\Delta$Y/$\Delta$Z = 1.5 with the primordial helium abundance Y$_p$ = 0.249. Once Y is computed for a value of Z, we calculate X as X = 1 -- Y -- Z. 

\textbf{Stellar rotation:} Rotation can significantly alter the evolution of massive stars \citep{heger00, heger05} and is  particularly important for our massive stellar models. We compute models both with and without rotation. Following the prescription described \citet{choi16}, we initialize our rotating stars to begin with solid body rotation at the Zero Age Main-Sequence (ZAMS). We use \vvcrit\ values of 0, 0.2, and 0.4, where \vvcrit\ is the critical surface linear velocity, defined at the equator of the star as:
\begin{equation}\label{eq:vcrit}
v^2_{\rm crit} = \left( 1 - \frac{L}{L_{\rm Edd}} \right) \frac{GM}{R^3} ,
\end{equation}
where the Eddington luminosity $L_{\rm Edd}$ is:
\begin{equation}
L_{\rm Edd} = \frac{4 \pi G M c}{\kappa},
\end{equation}
for a star with mass $M$, radius $R$, luminosity $L$, and opacity $\kappa$, and speed of light $c$. We adopt a fiducial rotation rate \vvcrit\ = 0.4 wherever the rotation rate is not explicitly mentioned as this value is frequently used as a standard rotating rate for models of massive stellar evolution and is supported by theoretical models of massive star formation independent of metallicity \citep{rosen12}. We do not implement rotation rates faster than \vvcrit\ = 0.4 as \citet{roy20} finds qualitatively similar results for all models with \vvcrit\ $> $ 0.4. The results of our grid of stellar evolution models are described in the following section.

\section{Results}\label{sec:results}
Our primary goal is to produce extensive grids of stellar evolutionary tracks for massive stars that cover a wide range in stellar masses, ages, evolutionary phases, and metallicities at non-solar scaled abundances for ease of comparison to solar scaled abundance stellar tracks. Our high mass models are terminated at the end of core carbon burning (C-burn) stage, the point at which the central $^{12}$C mass fraction drops below $10^{-4}$.

\subsection{Stellar Tracks}\label{sec:tracks}
\begin{figure*}
\includegraphics[width=7.1in]{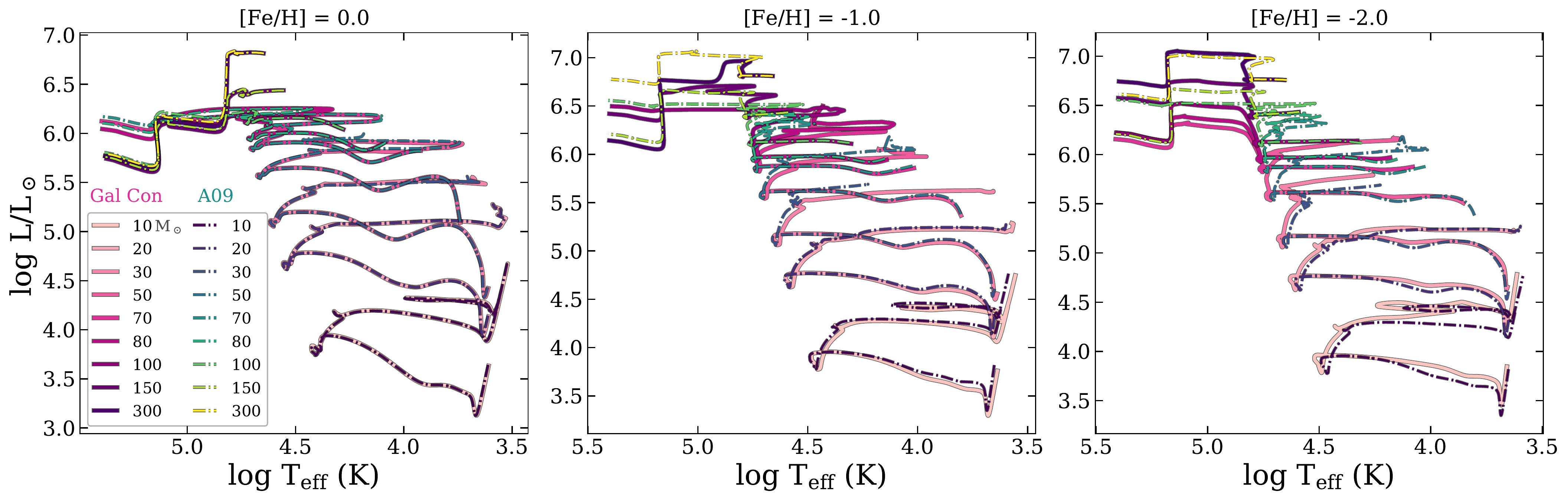}
\caption{[Fe/H] = 0.0 (left), [Fe/H] = $-$1.0 (middle), and [Fe/H] = $-$2.0 (right) grid of stellar evolutionary tracks over a wide range of stellar masses computed with \citet{asplund09} solar abundances (colored dotted lines) and Stromlo Galactic Concordance (Gal Con) abundances (solid pink lines) at different stellar masses. The stellar tracks for Galactic Concordance and Solar abundances are by design the same at [Fe/H] = 0.0 (left) and the tracks begin to deviate with decreasing metallicities due to non-uniform scaling of abundances with metallcities. }
\label{fig:HR_fehp0}
\end{figure*}

Figure~\ref{fig:HR_fehp0} shows MIST (solar) and Galactic Concordance evolutionary tracks at [Fe/H] = 0.0 and \vvcrit\ = 0.4. The differences in the tracks are relatively minor. More massive stars ($>150$~\Msun) with solar abundances tend to burn hotter and more luminous than the galactic concordance abundance tracks due to having a higher metal content. The change in the solar and non-solar scaled abundance tracks becomes more noticeable with increasing deviation at sub solar metallicities, a direct result of the non-uniform scaling of abundances with metallcities in the Stromlo tracks. This implies that the inferred ionizing spectra from these stars will have an increasingly important impact on the interpretation of HII region emission line spectra at metallicities less than solar. 

\begin{figure*}
\includegraphics[width=7.1in]{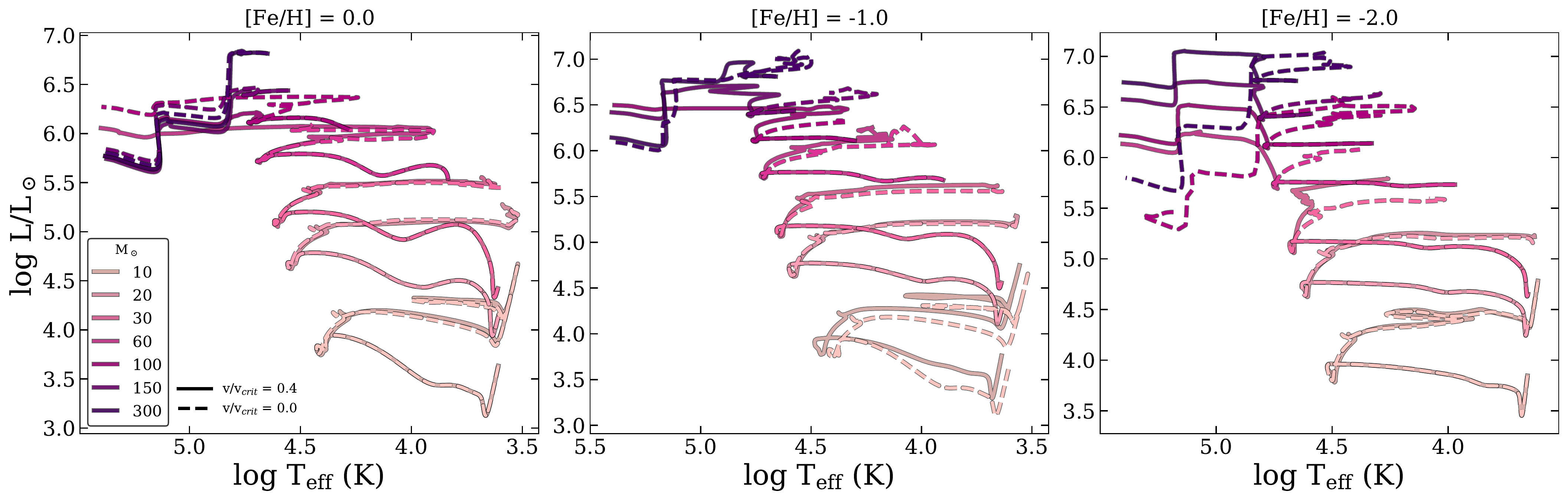}
\caption{[Fe/H] = 0.0 (left), [Fe/H] = $-$1.0 (middle), and [Fe/H] = $-$2.0 (right) grid of Stromlo stellar evolutionary tracks showing the effect of rotation on the evolutionary tracks computed with Galactic Concordance abundances at different stellar masses. Models with rotation are shown in solid lines with \vvcrit\ = 0.4 and non-rotating \vvcrit\ = 0.0 models are shown in dashed lines. Rotation plays a significant role with abundances (Figure~\ref{fig:HR_fehp0}) in determining $T_{\rm eff}$ and luminosity. }
\label{fig:HR_fehp0_rotating}
\end{figure*}

Figure~\ref{fig:HR_fehp0_rotating} shows the impact of rotation on the Galactic Concordance and solar stellar evolutionary tracks at [Fe/H] = 0.0, [Fe/H] = $-$1.0, and [Fe/H] = $-$2.0. The rotating models tend to be hotter and more luminous overall than the non-rotating star models as a result of the reduced mean opacity in rotating stars. The same effect in solar scaled MIST tracks was also reported in \citet{choi17}. The effect of rotation on the stellar evolutionary tracks has the largest impact at the lowest metallicities ([Fe/H] = $-$2.0 in Figure~\ref{fig:HR_fehp0_rotating}). Overall, rotation in stars has a larger impact than abundance ratio changes in the stellar tracks (Figure~\ref{fig:HR_fehp0}), with the differences most noticeable at the lowest metallicities ([Fe/H] = $-$2.0) as well.

\begin{figure*}
\includegraphics[width=7.1in]{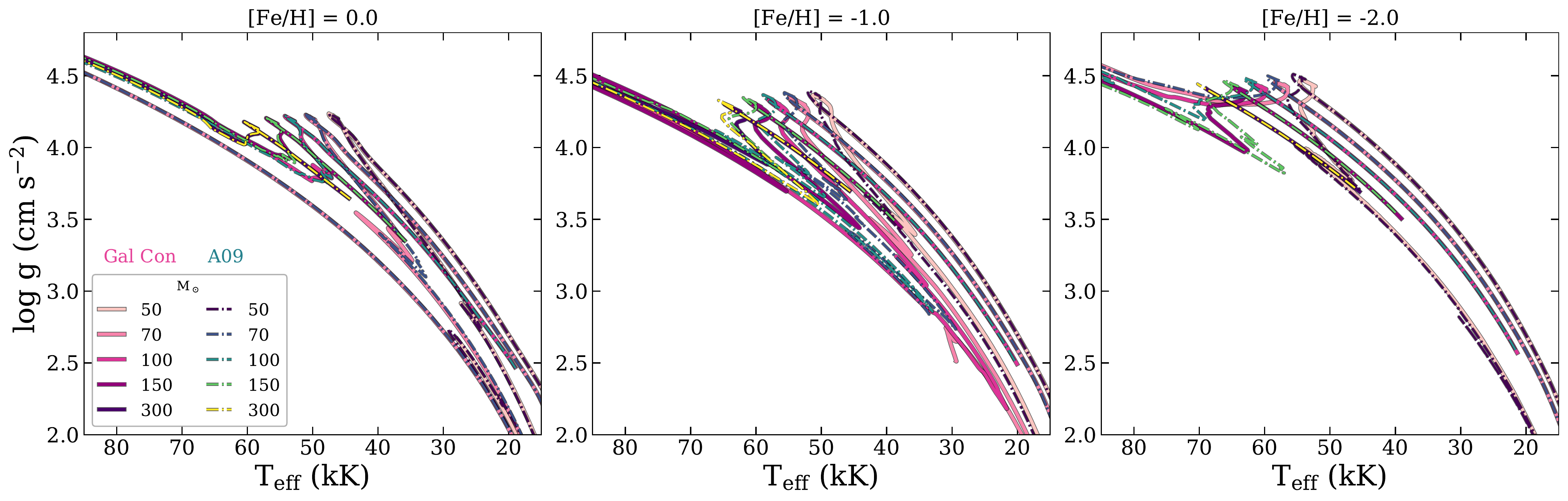}
\caption{[Fe/H] = 0.0 (left), [Fe/H] = $-$1.0 (middle), and [Fe/H] = $-$2.0 (right) grid of stellar evolutionary tracks showing the surface gravity as a function of effective temperatures over a range of stellar masses computed with \citet{asplund09} solar abundances (A09; colored dotted lines) and Stromlo Galactic Concordance abundances (Gal Con; solid pink lines) at different stellar masses. The stellar tracks for Galactic Concordance and Solar abundances are similar at Solar metallicity and start to deviate and increase in their relative importance with decreasing metallicities. }
\label{fig:g_fehp0}
\end{figure*}

\begin{figure*}
\includegraphics[width=7.1in]{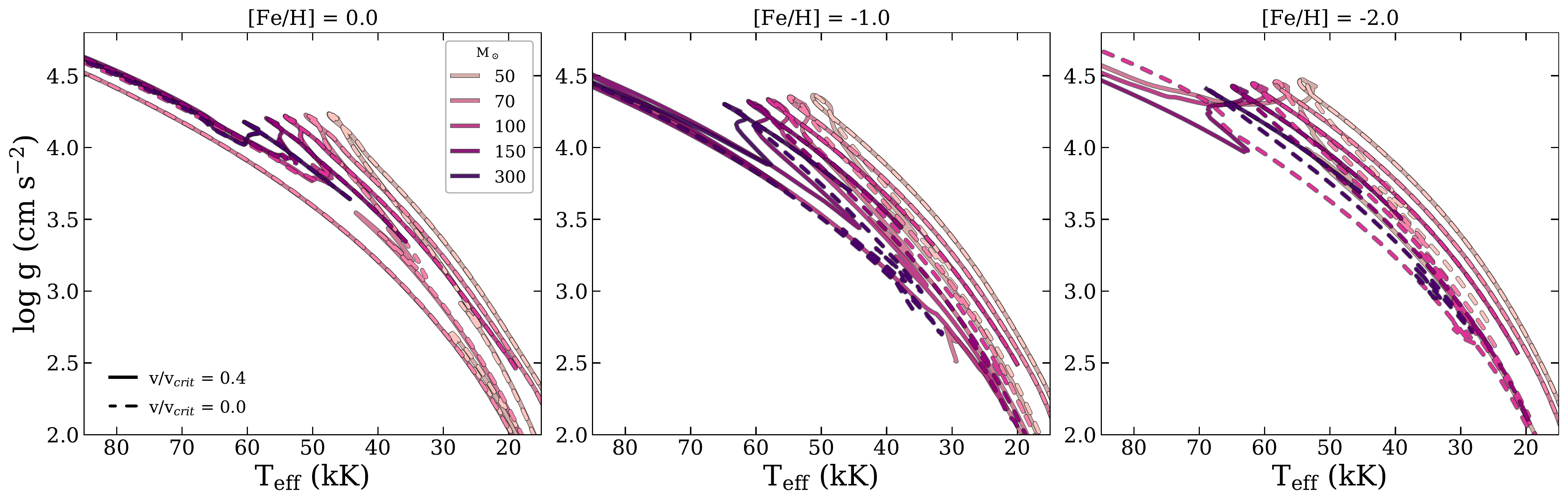}
\caption{[Fe/H] = 0.0 (left), [Fe/H] = $-$1.0 (middle), and [Fe/H] = $-$2.0 (right) grid of Stromlo stellar evolutionary tracks displaying the effect of rotation on the surface gravity as a function of effective temperatures for stellar tracks computed with Galactic Concordance abundances at different stellar masses. Models with rotation are shown in solid lines with \vvcrit\ = 0.4 and non-rotating \vvcrit\ = 0.0 models are shown in dashed lines. Rotation starts to play a more significant role than abundances alone (Figure~\ref{fig:g_fehp0}) determinations in determining the stellar properties at low ([Fe/H] $< -$1.0) metallicities. }
\label{fig:g_rotating}
\end{figure*}

Figures~\ref{fig:g_fehp0} and \ref{fig:g_rotating} show the stellar surface gravity $g$ as a function of the effective temperature $T_{\rm eff}$ and compares the impact of abundances and stellar rotation. Similar to a star's luminosity(Figure~\ref{fig:HR_fehp0}), we find that for stars with metallicities [Fe/H]$\gtrsim -1$, the galactic concordance abundance patterns have an equally important role as rotation (Figure~\ref{fig:g_rotating}) in determining the intrinsic stellar properties of $T_{\rm eff}$ and surface gravity $g$. For stars at low metalicities ([Fe/H] $=-2$), rotation has a larger impact in determining the stellar properties (e.g., $T_{\rm eff}$, surface gravity $g$, luminosity) than the correct abundance patterns. 

\subsection{Rotational Mixing}\label{sec:rotmixing}
In the previous section, we examined the impact that rotation plays in determining $T_{\rm eff}$ and luminosity, which appears to be as important as the chemical abundances. $T_{\rm eff}$ and luminosity are closely connected to mass-loss and surface abundances of different elements and we demonstrate here the measured surface abundance of our massive stars and its relation with rotational mixing. 

\begin{figure*}
\includegraphics[width=7.1in]{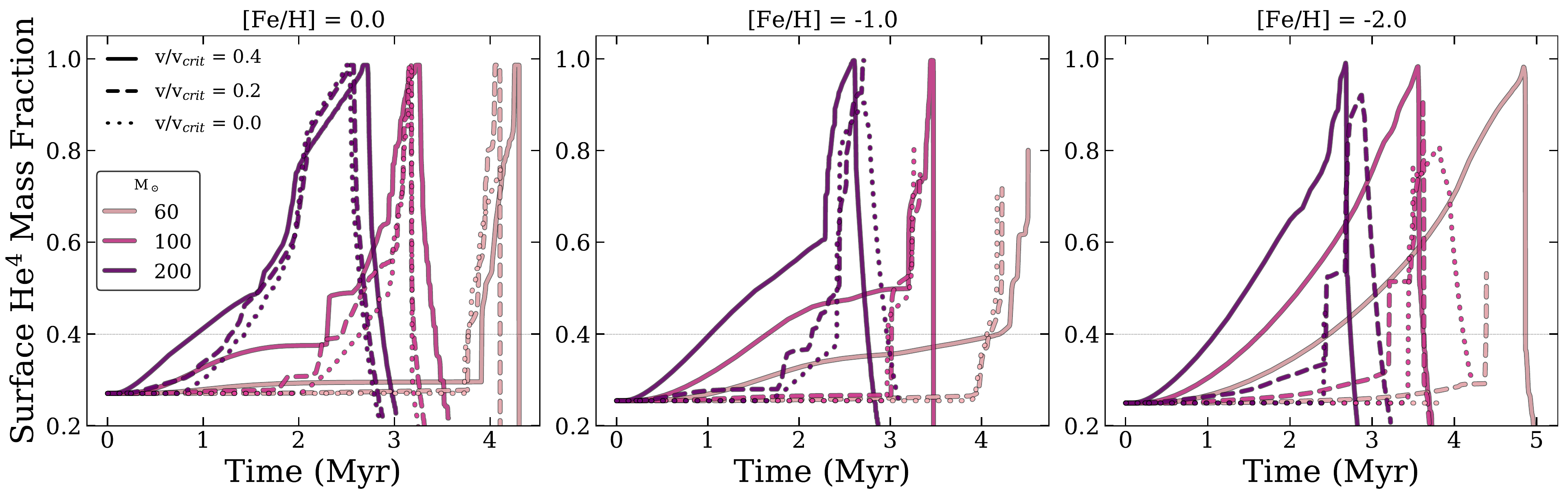}
\caption{Time evolution of the surface $^4$He abundances for 60, 100, and 200 \Msun\ stars with \vvcrit\ = 0.4 (solid lines), 0.2 (dash lines), and 0.0 (dot lines) at metallicities of [Fe/H] = 0.0 (left), [Fe/H] = $-$1.0 (middle), and [Fe/H] = $-$2.0 (right) for the Stromlo tracks. The gray line at 0.4 marks the He abundance that delineates the start of the WR phase of stellar evolution \citep{meynet05}.} 
\label{fig:surfaceabundace}
\end{figure*}

Figure~\ref{fig:surfaceabundace} shows the surface $^4$He mass fraction as a function of time for a sample of initial stellar masses, metallicities ([Fe/H] = 0.0, $-$1.0, and $-$2.0), and three rotation rates (\vvcrit = 0.4, 0.2, and 0.0). Figure~\ref{fig:surfaceabundace} demonstrates that rotation in stars more massive than $\sim100$~\Msun\ heavily impacts the surface composition; massive, rotating stars spend the vast majority of their lives on the main sequence with enhanced He surface abundances consistent with Wolfe Rayet stellar evolutionary phase \citep[e.g, ][]{abbott87, crowther95, meynet03, meynet05, roy20}. Rotational mixing and the enhancement of the surface He begins very fast; the surface He abundance rises to over 40\% just after $\sim$2~Myr for rotating \vvcrit = 0.4 stars of $\gtrsim$100~\Msun\ at metallicity [Fe/H] = 0.0. The two convective zones (inner core and outer core) of rotating stars are connected by three dominant rotational transport mechanisms: meridional circulation (ES circulation), GSF instability, and Spruit dynamo mixing \citep{roy20}. For non-rotating massive stars, there are no diffusion mechanisms for the transport of chemical elements from the inner convective core to the outer convective shell, and therefore there is no surface He enhancement. Rotational mixing enhances the surface composition of the massive stars with the effect increasing with decreasing metallicity. 

\begin{figure*}
\includegraphics[width=7.1in]{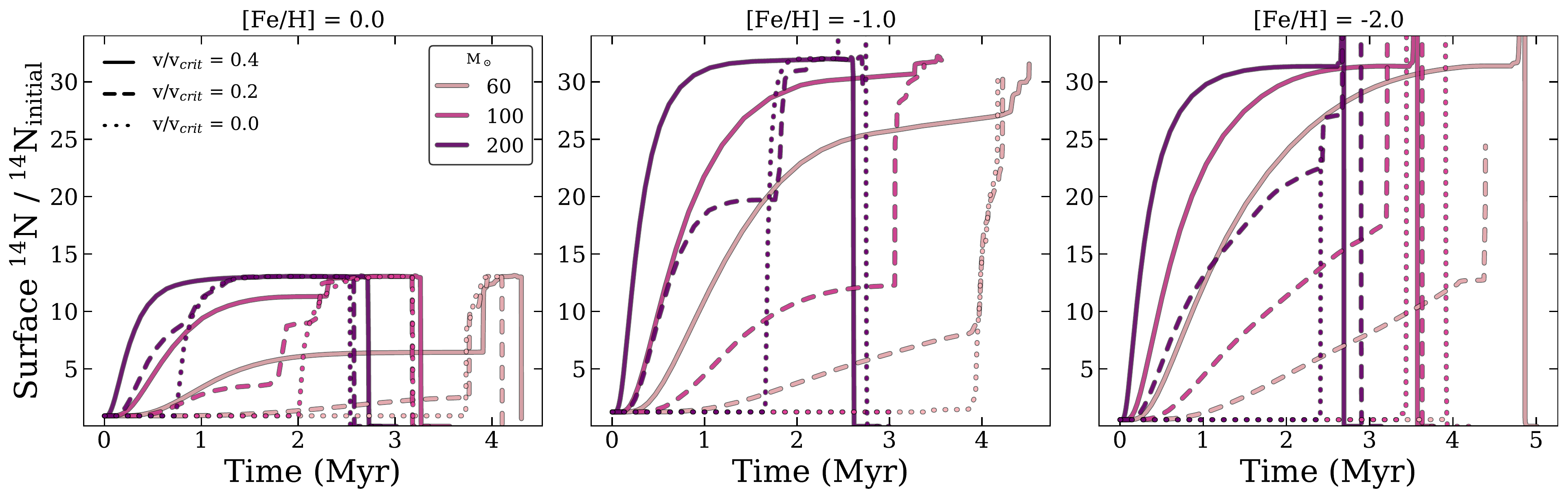}
\caption{
Time evolution of the surface $^{14}$N abundances normalizes by the initial $^{14}$N$_{\rm initial}$ abundance for 60, 100, and 200 \Msun\ stars with \vvcrit\ = 0.4 (solid lines), 0.2 (dash lines), and 0.0 (dot lines) at metallicities of [Fe/H] = 0.0 (left), [Fe/H] = $-$1.0 (middle), and [Fe/H] = $-$2.0 (right) for Stromlo tracks. The initial $^{14}$N$_{\rm initial}$ abundances are $6.11 \times 10^{-4}$, $8.70 \times 10^{-5}$, and $4.02 \times 10^{-6}$ for [Fe/H] = 0.0, [Fe/H] = $-$1.0, and [Fe/H] = $-$2.0, respectively. }
\label{fig:Nsurfaceabundace}
\end{figure*}

Figure~\ref{fig:Nsurfaceabundace} shows the enhancement of the surface mass fractions $^{14}$N normalized by the initial $^{14}$N$_{\rm initial}$ abundance, as a function of time and rotation rate for a range of initial stellar masses and metallicities. Figure~\ref{fig:Nsurfaceabundace} and Figure~\ref{fig:surfaceabundace} demonstrate that the surface mass fractions of helium and nitrogen are enhanced at low metallicity regardless of the rotation and enhanced at high rotation regardless of the metallicity. The enhancement is more prominent for more massive stars. At [Fe/H] = 0.0, our Galactic Concordance stars show a maximum nitrogen surface enhancement of $\sim14$. For the low metallicity stars [Fe/H] $<$--1.0, the nitrogen enhancement is a factor of $\sim$30--32 relative to their initial $^{14}$N abundances. 

The effect of rotational mixing on the observed surface abundance chemical elements have been noted before \citep[][]{heger00, meynet05, crowther07} and will impact the observed surface opacity-age relationships. This in turn will impact the modeled ionizing photon spectrum output compared to stars with non-enhanced He, N surface compositions \citep[e.g., ][]{choi17, roy20}. 

\subsection{Main sequence lifetimes}\label{sec:mslifetimes}
\begin{figure*}
\includegraphics[width=7.1in]{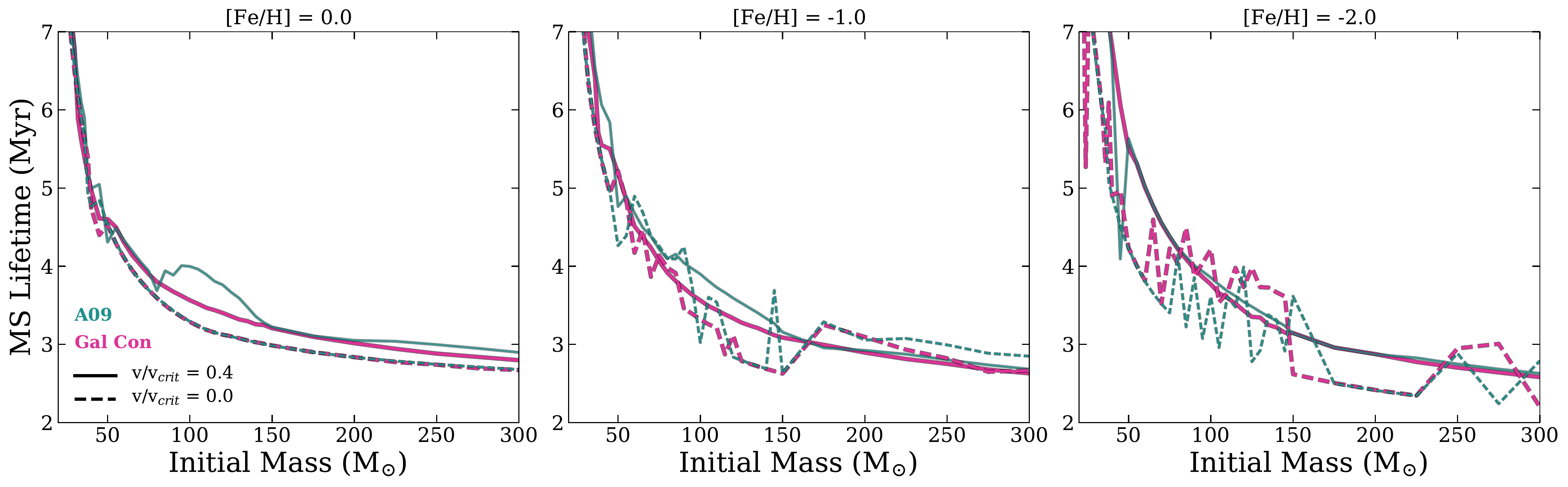}
\caption{Main sequence lifetimes as a function of initial mass for two different values of initial rotation rates (\vvcrit\ = 0.4 in solid lines and \vvcrit\ = 0.0 in dotted lines) at [Fe/H] = 0.0 (left), [Fe/H] = $-$1.0 (middle), and [Fe/H] = $-$2.0 (right). Our new Stromlo tracks with Galactic Concordance (Gal Con) abundances are shown in pink and the MIST tracks \citep{choi16} are shown in green. At a fixed initial mass, higher rotation rates lengthen the main sequence lifetime due to more efficient rotational mixing. The jagged nature of the lines for decreasing metallicity highlights convergence issues for models that do not run to completion. 
} 
\label{fig:MSlifetime_fehp0}
\end{figure*}

The enhanced mixing in the cores of rotating stars responsible for increasing their brightness also extends the main sequence lifetimes of the stars. Figure~\ref{fig:MSlifetime_fehp0} shows the main sequence lifetime--initial mass relation. As expected, the main sequence lifetime is longer for rotating stars due to rotational mixing channeling additional fuel into the core of the star. The main sequence lifetimes for the Stromlo rotating models and MIST agree to within 10\% at solar metallicity, though the main sequence lifetimes are shorter for the Stromlo tracks at solar metallicity.  Our rotating Galactic Concordance tracks at solar metallicity do not show the non-monotonic behavior at 80~\Msun\ with prolonged main sequence lifetimes as seen in the \vvcrit\ = 0.4, [Fe/H] = 0.0 metallicity MIST tracks. For all rotating stars, the boost to the main sequence lifetime at a fixed initial mass is marginally larger for the MIST tracks than for the Stromlo tracks, suggesting that rotational mixing may be marginally more efficient in solar-scaled stars. 

\subsection{Stellar Isochrones}\label{sec:isochrones}
We note that in any grid, there are subsets of models that do not run to completion due to convergence issues \citep[][Figure~\ref{fig:MSlifetime_fehp0}]{choi17}. In general this is not an issue because the mass sampling is sufficiently fine that there are enough models to smoothly interpolate and construct smooth isochrones with the tracks that are available using the same mass grid as MIST \citep{dotter16}. We do note that at low metallicity, the mass grid sampling as laid out by MIST is not always sufficient to represent the fast evolutionary phases at early times, however, we do not model ancient, metal-poor populations in this current paper ([Fe/H]=$-4$) as is done by \citet{choi17} due to computational and convergence difficulties at these very low metallicities. The isochrones for our Stromlo stellar tracks are computed at three rotation rates \vvcrit\ = 0.0, 0.2, 0.4 and at each metallicity point in our grid. Because we only compute the stellar evolutionary tracks from 10--300~\Msun, our isochrones are only valid for stars $\lesssim$25~Myr old. The Stromlo and solar stellar evolutionary tracks are processed into isochrones following the procedure outlined in \citet{dotter16}. 

\begin{figure*}
\includegraphics[width=7.1in]{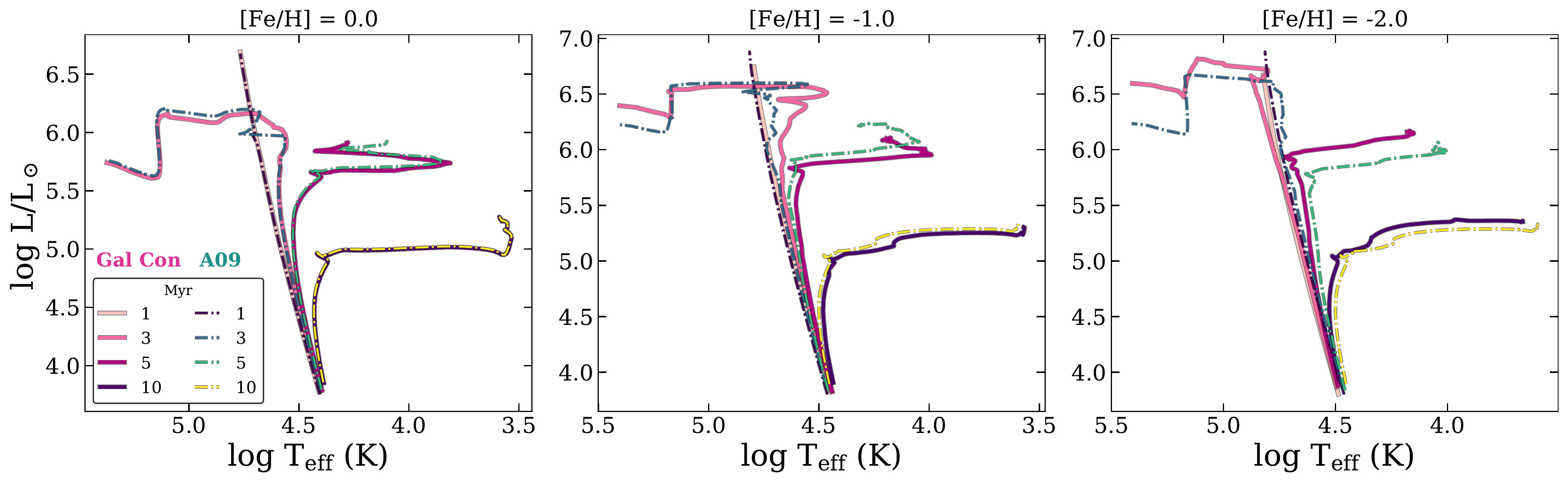}
\caption{[Fe/H] = 0.0 (left), [Fe/H] = $-$1.0 (middle), and [Fe/H] = $-$2.0 (right) isochrones for solar (A09; dotted colored lines) and Stromlo Galactic Conordance (Gal Con; pink solid lines) abundances as a function of age for rotation rate \vvcrit\ = 0.4. The four colors correspond to ages 1, 3, 5, and 10~Myr, respectively. At solar metallicity, solar abundance and Galactic Concordance tracks show similar results. At very low metallicities [Fe/H] = $-$2.0, Galactic Concordance tracks are in general more luminous. 
} 
\label{fig:iso_fehp0_gc_a09}
\end{figure*}

\begin{figure*}
\includegraphics[width=7.1in]{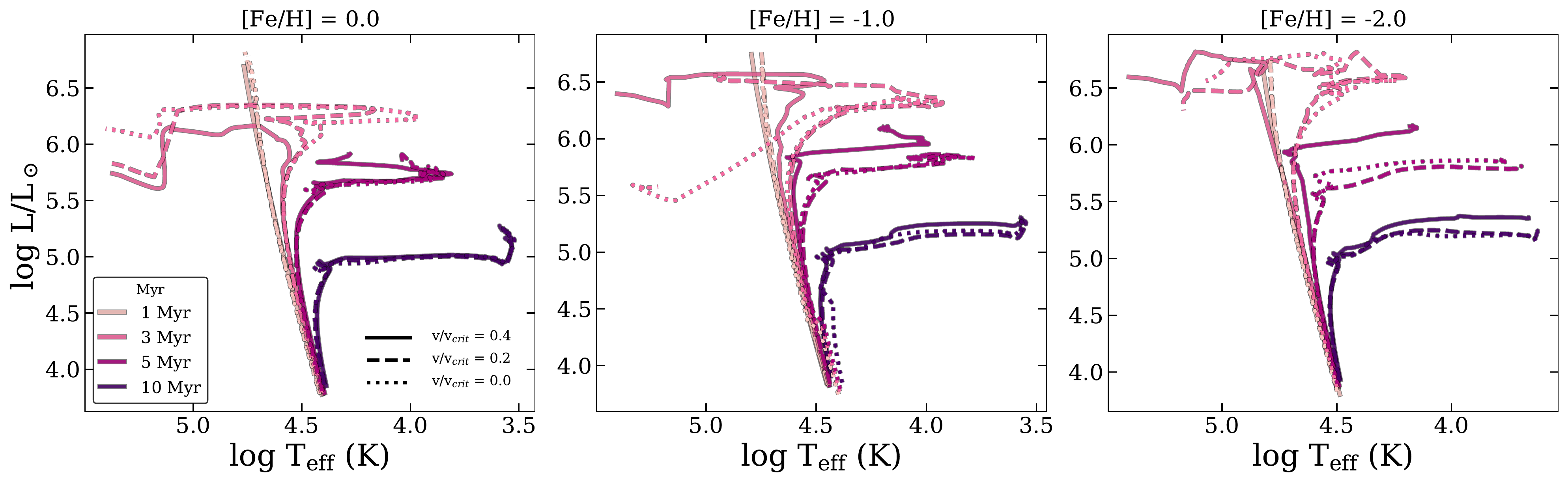}
\caption{[Fe/H] = 0.0 (left), [Fe/H] = $-$1.0 (middle), and [Fe/H] = $-$2.0 (right) Stromlo isochrones as a function of age and rotation. The four colors correspond to ages 1, 3, 5, and 10~Myr, respectively. The linestyle refers to the rotation of the star. Fast rotation generally leads to hotter, brighter, and longer-lived stars. The rotation of the star is significant in determining the luminosity and $T_{\rm eff}$ as the initial abundances of the stars (Figure~\ref{fig:iso_fehp0_gc_a09}). 
} 
\label{fig:iso_fehp0_multirotation}
\end{figure*}

Figures~\ref{fig:iso_fehp0_gc_a09} and \ref{fig:iso_fehp0_multirotation} show 1, 3, 5, and 10~Myr isochrones at [Fe/H] = 0.0 and [Fe/H] = $-$1.0 and at varying rotations. The effect of faster rotation resulting in hotter, brighter, and longer-lived stars is immediately clear. The Stromlo tracks with Galactic Concordance abundances show the same fast appearance of Wolf-Rayet (WR) stars \citep[T$_{\rm eff} > 10^4$~K and surface hydrogen mass fraction $X < 0.3$;][]{meynet03, georgy12} from the massive star progenitors between 3--5~Myr (Figure~\ref{fig:surfaceabundace}), also observed in the massive rotating stars with solar abundances \citep{choi17}. 

\section{Discussion}\label{sec:discussion}
In this section we discuss the implications of rotating, massive star models with non-Solar elemental abundances on the interpretation of high-redshift star-forming galaxies. We examine applications for physical conditions at high and low redsihft where these hot, rotating star models will have a dramatic impact.

Abundance scaling of elements at metallicities lower than the Solar standard has been explored in stars, but in the nebular modelling community, only simple uniform scaling assumptions are typically implemented. The surface enhancement of elements in massive rotating stars will have broad implication for the ionizing spectra of high-redshift, low-metallicity galaxies. Under the assumption that stars retain their original surface composition until they leave the main sequence, there is minimal impact on the hydrogen-ionizing photon fluxes, and thus the effect of this assumption on the total ionizing photon budget is minimal. Moderate and rapidly rotating stars heavily impact the original surface composition of these massive stars and while the hydrogen ionizing flux remains relatively unchanged, the self-consistently evolved surface composition for fast rotating stars contains fewer photons at the helium ionizing edge \citep{roy20}. The different surface opacity-age relationships make significant contribution to the photon ionizing budget than would be expected if the surface composition were un-evolved. This increases the complexity in interpreting spectral line diagnostics and measuring reliable metallicity indicators. Consequently, stellar models in conjunction with photoionization models for nebulae with different metallicities need to take this variation into account in order to accurately model and predict the properties of galaxies across cosmic time. 

To investigate the impact on the ionizing spectrum of our stellar evolutionary tracks computed using non-Solar scaled abundances, we model the simple stellar populations (SSPs) of our stellar isochrones (Section~\ref{sec:isochrones}) using the Flexible Stellar Population Synthesis package \citep[FSPS;][]{conroy09, conroy10}. We predict the simulated SED for a 10$^6$~\Msun\ stellar population following an instantaneous burst of star formation and a fully sampled \citet{kroupa01} IMF with limits of 0.08 -- 300~\Msun. We implement the MILES empirical library as the primary stellar spectral library \citep{sanchez-blazquez06, falcon-barroso11}. 

\begin{figure*}
\includegraphics[width=7.1in]{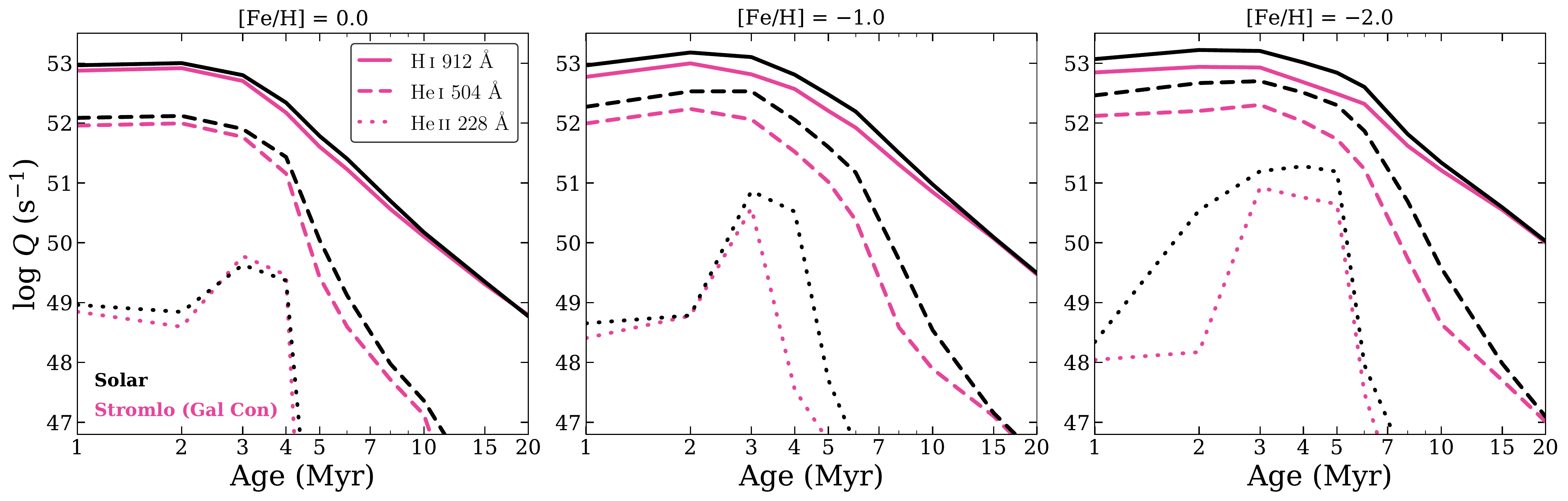}
\caption{Time evolution of the ionizing photon luminosity $Q$ for the Stromlo models (Gal Con; pink) and models assuming Solar abundances (black) for a 10$^6$~\Msun\ stellar population at [Fe/H] = 0.0 (left), $-$1.0 (middle), and $-$2.0 (right). The linestyle represent the ionizing photons capable of ionizing hydrogen H\,{\textsc{i}} (912~\AA; solid lines), singly ionizing helium He\,{\textsc{i}} (504~\AA; long dashed lines), and doubly ionizing helium He\,{\textsc{ii}} (228~\AA; short dotted lines). The ionizing photon output for solar versus galactic concordance abundances show broad agreement with the galactic concordance stars showing a softer spectrum with the differences becoming more pronounced at lower metallicities.  
} 
\label{fig:sed}
\end{figure*}

We compute the time evolution of the ionizing photon luminosity $Q$ from 1 to 20~Myr for 10$^6$~\Msun\ stellar population at [Fe/H] = $-$2.0, $-$1.0, and 0.0, shown in Figure~\ref{fig:sed} computed at \vvcrit\ = 0.4 for stellar models with Soloar and Galactic Concordance abundances. The models produce comparable hydrogen and helium-ionizing photon output rates at [Fe/H] = 0.0. The difference of elemental abundances on the resulting ionizing photon output begins to deviate more at lower metallicity with the impact occurring at the highest energies blueward of 228~\AA. The Stromlo models predicts a soften spectrum than the Solar models, primarily due to differences in the stellar populations of the underlying stellar models (Figure~\ref{fig:HR_fehp0}). The Stromlo models cease to produce an appreciable amount of photons blueward of 228~\AA\ (the wavelength of photons capable of doubly ionizing helium) beyond 5~Myr, 1~Myr shorter than what is seen for the Solar models. Single-star models rely exclusively on the most massive stars as the principal ionizing sources and the ionizing photon output decreases dramatically upon the disappearance of the most massive stars the first few Myr. 

\begin{figure}
\includegraphics[width=3.4in]{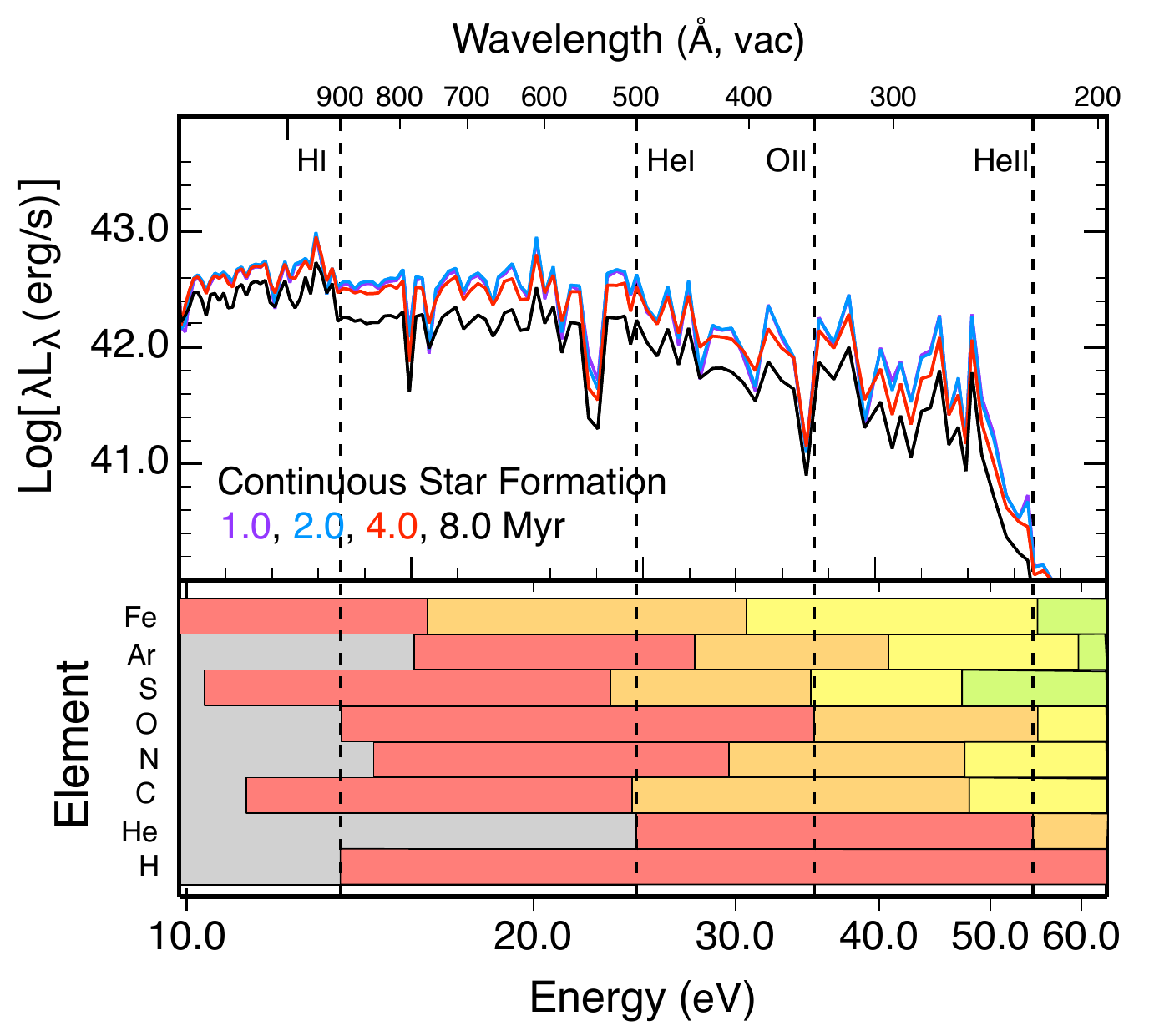}
\caption{The spectral energy distribution for FSPS simulations of star clusters with ages of 1$-$8~Myr (different colored lines) computed with a continuous star formation. The dotted black lines show the energy (wavelength on top axis) for singly ionized hydrogen H\,{\textsc{i}}, singly ionized helium He\,{\textsc{i}}, doubly ionized oxygen O\,{\textsc{ii}}, and doubly ionized helium He\,{\textsc{ii}}. The stellar spectra have been normalized to $\log L_\lambda = 40$~erg/s/\AA\ at 912~\AA. The bottom panel shows select ions and the corresponding energy bands where gray is neutral, red is the region where the ions are singly ionised, orange is doubly ionized, yellow is triply ionized, and green is fourthly ionized. 
} 
\label{fig:sb99}
\end{figure}

It is important to stress that the only changes made to the stellar tracks to compute the ionizing luminosity $Q$ in Figure~\ref{fig:sed} is the relative elemental abundances from Solar to Galactic Concordance for the Stromlo tracks. The differences in the the singly ionizing helium He\,{\textsc{i}} (24.6~eV) ionizing flux indicate this change to the tracks could affect important nebula lines compared to tracks computed with solar abundances. There are a large number of key nebular lines between the energy necessary to singly ionize helium (24.6~eV) and the energy required to doubly ionize helium (54.4~eV; Figure~\ref{fig:sb99}). This indicates that small changes in the stellar tracks between singly and doubly ionized helium will potentially have a large impact in the nebula ionization and subsequent nebula lines important for local metallicity and star formation rates as well as predictions for photoionization calculations in the early universe. 

\begin{figure*}
\includegraphics[width=7.1in]{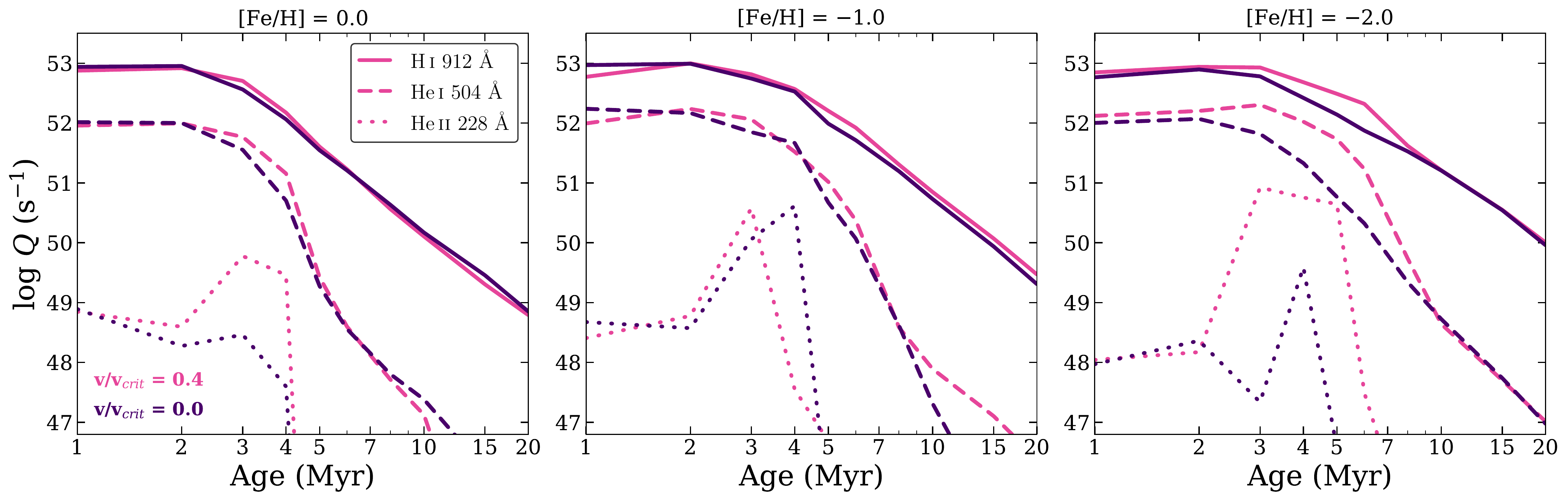}
\caption{Time evolution of the ionizing photon luminosity $Q$ for the Stromlo models for a 10$^6$~\Msun\ stellar population at [Fe/H] = 0.0 (left), $-$1.0 (middle), and $-$2.0 (right) for two different values of initial rotation rates (\vvcrit\ = 0.4 in pink lines and \vvcrit\ = 0.0 in dark purple lines). The linestyle represent the ionizing photons capable of ionizing hydrogen H\,{\textsc{i}} (912~\AA; solid lines), singly ionizing helium He\,{\textsc{i}} (504~\AA; long dashed lines), and doubly ionizing helium He\,{\textsc{ii}} (228~\AA; short dotted lines). Non-rotating stars show a softer spectrum than rotating stars and the differences become more pronounced at lower metallicities and is most significant at the highest energies shortward of 228~\AA.  
} 
\label{fig:sed_rotating}
\end{figure*}

Figure~\ref{fig:sed_rotating} shows the impact of stellar rotation of the Stromlo models on the ionizing photon output. The rotating and non-rotating stars show broad agreement with each other; discrepancies between the (non)rotating models are only significant at the highest energies shortward of 228~\AA. The minimal impact of the stellar rotation on the hydrogen ionizing luminosity compared to the helium ionizing luminosity has been demonstrated in prior studies \citep{levesque12, choi17}. The galactic concordance abundance patterns have an equally important role as rotation in determining the ionizing photon output from the stellar properties, especially at lower metallicities; adopting stellar models that are not solar-scaled are critical ingredients in modeled SEDS and the interpretation of observations using these models. Considering non-Solar scaled abundances in stellar models is critical for accurate measurements of the ionizing photon budget from star clusters in local studies as well as measurements for the escape fraction of ionizing photons in high-redshift galaxies during cosmic reionization.

Constraining the impact of non-Solar scale elemental abundance self-consistently with atmosphere and nebular modeling and its importance for spectral line diagnostics await a full spectral synthesis calculation in future work. This paper is the first in a series to approach this problem in a self-consistent manner and additionally requires a large library of atmospheres covering the full range of physical properties and atmospheric compositions. The atmospheric library and and their application to complete nebular radiative transfer modeling will be investigated in the future (R. Sutherland et al., in prep). A future paper will investigate the variable non-Solar elemental abundances stellar tracks and how these abundance composition changes affect the modeled spectrum and quantify the impact on the ionizing photon budget and spectral line diagnostics (K. Grasha et al., in prep) and the implication of elemental abundances of massive star models for cosmic reionization and the interpretation of high-redshift star-forming galaxies.

\section{Summary and Conclusions}\label{sec:conclusions}
In this paper we investigate the impact of the stellar elemental abundances on the stellar evolutionary tracks in massive rotating and non-rotating stars over a wide range of metallicities using the Galactic Concordance abundances from resolved Milky Way \HII\ regions. We use MESA for our stellar evolutionary calculations and include all the same physical processes and parameters values as used in the MIST track library \citep{choi16} for uniform scaled solar-abundances. We focus on massive stars evolved until the end of the main-sequence phases and adopt the same the modifications to improve the treatment of massive stars and the implementation of Galactic Concordance abundances as outlined by \citet{roy20}.

We summarize our main conclusions below. 

\begin{enumerate}

\item The assumed elemental abundance ratios have minor influence on massive stellar evolutionary tracks at solar metallicities (Figure~\ref{fig:HR_fehp0}). The correct implementation of non-uniformly scaled elemental abundance ratios becomes more significant at sub-solar metallicities, where the differences between Stromlo stars and MIST uniform solar-scaled abundance stars are primarily most pronounced for stars more massive than 50~\Msun at low metallicities. 

\item Rotation and abundances both play a significant role in determining the stellar parameters of $T_{\rm eff}$, luminosity, and surface gravity $g$ (Figure~\ref{fig:HR_fehp0_rotating}, Figure~\ref{fig:g_fehp0}). Rotation effects are more significant at lower metallicities in determining $T_{\rm eff}$, luminosity, and surface gravity $g$ than abundance patterns as the stars become more compact and angular momentum loss due to winds becomes less important. 

\item The significant effect of rotation in determining the stellar $T_{\rm eff}$, luminosity and surface gravity $g$, especially at low metallicities, has a large impact on mass-loss and surface abundances of different elements (Figures~\ref{fig:surfaceabundace} and \ref{fig:Nsurfaceabundace}). Rapidly rotating stars (\vvcrit = 0.4) more massive than 100~\Msun\ show helium surface abundance enhancement to 40\% by $\sim2$~Myr. Even in non-rotating stars, toward the end of the core H depletion, stars experience the `classical' WR phase, approaching the Eddington limit and experiencing rapid mass-loss, which enhances the He surface abundance upwards to 90\%. At lower metallicities ([Fe/H] = $-2$), the effect of rotation on He surface abundance enhancement becomes more significant with He surface enhancement occurring as rapidly as 1.5~Myr for 100~\Msun\ stars. Rotation plays an extremely important role in the surface enhancement of nitrogen, with the effect becoming more important at lower metallicities. 

\item Rotation lengthens the main sequence timescale of massive stars (Figure~\ref{fig:MSlifetime_fehp0}) and leads to brighter and hotter stars (Figure~\ref{fig:iso_fehp0_multirotation}). The boost in the main sequence lifetime at a fixed mass is slight larger in our Galactic Concordance Stromlo tracks than in the MIST solar-scaled models in both rotating and non-rotating stars at low metallicities ([Fe/H] = $-2$; Figure~\ref{fig:iso_fehp0_gc_a09}), a result of increased metallicity and a decrease in the mean opacity. 

\item The Stromlo tracks show a softer ionizing spectrum compared to expectations from the Solar-scaled ionizing spectrum (Figure~\ref{fig:sed}). The ionizing photon luminosity between Solar and Galactic Concordance models deviate most significantly at low metallicities. The stellar populations in low metallicity environments, common at high redshift, only require moderate rotation rates to produce significant ionizing photons, decreasing rapidly after the disappearance of the most massive stars after a few Myr.

\end{enumerate}

This paper is the first implementation of Galactic Concordance abundances to the stellar evolution models that does not implement solar, scaled-solar, or alpha-element enhanced abundances. These models will be applicable for extreme regions of star formation, especially low metallicity systems and active star-forming galaxies, where massive and rotating star models at non-uniform scaled metallicitices have the potential to heavily impact the resulting properties of the star-forming emission line regions. The importance of rotating, massive star models and their elemental abundance scaling have broad implications within the context of cosmic reionization and the interpretation of high-redshift star-forming galaxies. 

The Galactic Concordance scale is not the sole scaling parameter that can be used for non-uniform scaling relations between elements and metallicity. It is vital, however, that a single scale is implemented for all the necessary components involved within stellar population synthesis and photoionization models. In the future, we will investigate the effect of Galactic Concordance scaled abundances of massive, rotating stars in low metallicity environments using stellar population synthesis applications using self-consistent atmosphere and nebular modeling with MAPPINGS to quantify the impact to the ionizing photon budget and spectral line diagnostics.

\acknowledgements
We are grateful for the valuable comments on this work by an anonymous referee that improved the scientific outcome and quality of the paper.
This research was conducted on Ngunnawal Indigenous land. 
KG gratefully acknowledges the support of Lisa Kewley’s ARC Laureate Fellowship (FL150100113).
AR acknowledges the usage of Australian National University RSAA cluster AVATAR and NCI GADI via project jh2 for the implementation of the Galactic Concordance (and also any arbitrary) abundance setups for this work. Also, AR gratefully acknowledges the support of Mark Krumholz's Discovery Project (DP160100695) and Future Fellowship (FT180100375) award grants. 
This research has made use of NASA's Astrophysics Data System Bibliographic Services. This research made use of Astropy,\footnote{http://www.astropy.org} a community-developed core Python package for Astronomy \citep{astropy13, astropy18}.
Parts of this research were supported by the Australian Research Council Centre of Excellence for All Sky Astrophysics in 3 Dimensions (ASTRO 3D), through project number CE170100013. 
The authors thank the invaluable labor of the maintenance and clerical staff at their institutions, whose contributions make scientific discoveries a reality.

\software{Astropy \citep{astropy13, astropy18}, iPython \citep{ipython}, Matplotlib \citep{matplotlib}, Numpy \citep{numpy11, numpy20}, scipy \citep{scipy}, MESA \citep{paxton11}.}

\bibliography{ms_stromlotracks_submitted_arxiv2.bbl}

\end{document}